\documentclass[sigconf]{acmart}
\AtBeginDocument{%
  }

\setcopyright{acmlicensed}
\copyrightyear{2026}
\acmYear{2026}
\acmConference[ICSE 2026]{International Conference on Software Engineering - IEEE SWEBOK Summit 2026}{April 18,
  2026}{Rio de Janeiro, Brazil}
\begin{document}

\title{A Semantic-Web Oriented Competency Model for Engineering Programs}

\author{Nicolas Evain}
\email{nicolas.evain@univ-pau.fr}
\orcid{0009-0009-0730-3833}
\affiliation{%
  \institution{Universite de Pau et des Pays de l'Adour, E2S UPPA, LIUPPA}
  \city{Anglet}
  \state{Nouvelle-Aquitaine}
  \country{France}
}

\author{Ernesto Exposito}
\email{ernesto.exposito@univ-pau.fr}
\orcid{0000-0002-3543-2909}
\affiliation{%
  \institution{Universite de Pau et des Pays de l'Adour, E2S UPPA, LIUPPA}
  \city{Anglet}
  \state{Nouvelle-Aquitaine}
  \country{France}
}

\author{Philippe Arnould}
\email{philippe.arnould@univ-pau.fr}
\orcid{0000-0002-6850-2096}
\affiliation{%
  \institution{Universite de Pau et des Pays de l'Adour, E2S UPPA, LIUPPA}
  \city{Anglet}
  \state{Nouvelle-Aquitaine}
  \country{France}
}

\renewcommand{\shortauthors}{Evain et al.}

\begin{abstract}
Despite comprehensive Bodies of Knowledge (BoKs) documenting core knowledge across software engineering, computer science, information systems, and emerging computing fields, a critical gap persists: methodologies for integrating this knowledge into coherent competency-based curricula that prepare graduates for professional careers remain underdeveloped. This paper presents a competency-mapping methodology that bridges Bodies of Knowledge and competency frameworks to design computing curricula. We demonstrate this methodology through ISANUM, a five-year engineering degree program featuring 23 competencies organized into five thematic blocks, each with explicit mappings to 494 knowledge topics from 34 Computing Knowledge areas defined in Computing Curricula 2020. The program integrates three specialized pathways (Software Engineering, Data Engineering \& Data Science, and Information Technology) with mandatory work-study programs, ensuring graduates develop both theoretical foundations and practical workplace competencies. Our contribution provides computing educators with a replicable methodology for translating Bodies of Knowledge into assessable competency frameworks, supported by a semantic wiki infrastructure (ISANUMpedia) enabling collaborative curriculum understanding, maintenance and evolution.
\end{abstract}


\received{3 October 2025}
\received[accepted]{3 December 2025}

\maketitle
\section{Introduction}\label{sec1}

The computing field is characterized by rapid and continuous evolution, making it crucial to provide learners with fundamental knowledge that remains relevant despite technological changes~\cite{CC2020}. Every profession is based on a Body of Knowledge (BoK) which is defined as a collection of knowledge items or areas generally agreed to be essential to understanding a particular subject~\cite{SWEBOKV3}. In computing education, BoKs serve as source references that document the core knowledge fundamentals across various domains, including computer engineering, computer science, cybersecurity, information systems, information technology, software engineering, and data science~\cite{CC2020}.

Computing is a vast and complex field experiencing constant significant changes: societal needs evolve with computing becoming ubiquitous; technologies advance exponentially in storage capacity, processing power, communication, and interaction capabilities; and the methods by which computing professionals exploit resources evolve rapidly through new programming paradigms, development methodologies, and environments. This raises fundamental questions: what should be learned, what should be taught, and what is truly important when the computing landscape of today will differ significantly in 5, 10, or 20 years? Despite the availability of comprehensive BoKs, a critical gap exists in their integration into curriculum design, hindering the effective preparation of learners to address the digital transformation challenges faced by our modern society~\cite{CS2023}.

Existing computing curricula guidelines provide frameworks for educational program development~\cite{CC2020,CS2023,IS2020,IT2017,SE2014,CSEC2017}. However, research indicates limited alignment between these BoKs, curriculum implementations, and the competencies required in professional practice. Recent studies highlight significant misalignments between computing education and industry needs~\cite{adams2020_computing_dominate_stem}. The U.S. Bureau of Labor Statistics projects 165,000 annual computing job openings~\cite{bls_emp_by_detailed_occupation}. Despite these workforce demands, methodologies for translating BoKs into competency frameworks that address employment requirements remain underdeveloped.

This paper presents a competency-mapping methodology that leverages international standards and Bodies of Knowledge to design computing curricula aligned with competency development, course outcomes, and professional requirements. We demonstrate how to establish explicit mappings between knowledge areas defined in BoKs, competency frameworks from international standards, and learning outcomes in educational programs and structured learning pathways.

We present ISANUM, a five-year engineering degree program that implements this competency-mapping methodology through a competency-based framework. The program demonstrates how to bridge the gap between referential BoKs and practical workforce preparation through: (1) a competency model with 23 clearly defined competencies organized into five blocks, each with explicit mappings to BoK knowledge areas; (2) detailed knowledge maps that translate 494 BoK topics from 34 Computing Knowledge areas into teachable content; (3) three specialized pathways (Software Engineering, Data Engineering, and Information Technology) that address digital transformation challenges across multiple sectors; and (4) mandatory alternance, ensuring students develop both theoretical foundations and practical workplace competencies through work-integrated learning. This work provides computing educators with a replicable methodology for designing programs that connect Bodies of Knowledge to competency-based curricula, validated through the successful implementation of a French engineering degree program.
\vspace{-0.5em}
\section{Background: Bodies of Knowledge and Computing Curricula}\label{sec2}

The computing education landscape is structured around comprehensive Bodies of Knowledge (BoKs) and curricula recommendations that define the foundational knowledge required across various computing disciplines. The Computing Curricula 2020~\cite{CC2020} provides a contemporary, unified vision of computing education, identifying competencies and knowledge areas common across all computing disciplines. Each computing discipline maintains specialized Bodies of Knowledge and curriculum guidelines that document the evolving knowledge requirements for their respective domains:

\textbf{Software Engineering:} the software engineering discipline is documented through multiple authoritative sources. The Software Engineering Body of Knowledge~\cite{SWEBOKV4,SWEBOKV3} provides comprehensive coverage of software engineering knowledge areas, complemented by curriculum guidelines including Curriculum Guidelines for Undergraduate and Graduate Degree Programs in Software Engineering ~\cite{SE2014,SE2009,SE2004}. These documents establish the theoretical foundations and practical competencies essential for software engineering professionals.

\textbf{Computer Science:} computer science education has evolved through successive curriculum recommendations, including the most recent Computer Science Curricula 2023~\cite{CS2023}, building upon previous versions CS2013~\cite{CS2013} and CS2008~\cite{CS2008}. These guidelines define core knowledge areas spanning algorithmic foundations, programming languages, systems fundamentals, and theoretical computer science.

\textbf{Information Systems:} information systems education integrates technical and organizational perspectives, documented in the Guide to the Systems Engineering Body of Knowledge (SEBoK)~\cite{SEBoK2024}, the IS2020 competency model~\cite{IS2020}, the MSIS2016 graduate program guidelines~\cite{MSIS2016}, and the IS2010 undergraduate curriculum~\cite{IS2010}. These resources emphasize the strategic role of information systems in organizational contexts.

\textbf{Information Technology:} information technology curricula, including IT2017~\cite{IT2017} and IT2008~\cite{IT2008}, focus on the application and integration of computing technologies. The Enterprise Information Technology Body of Knowledge (EITBOK)~\cite{eitbok} complements these curriculum guidelines by documenting enterprise-scale IT practices and competencies.

\textbf{Computer Engineering:} computer engineering education bridges hardware and software domains, as defined in Computer Engineering Curricula 2016~\cite{CE2016}, Digital Design and Computer Architecture~\cite{DDCA2015}, and Computer Engineering Curricula 2004~\cite{CE2004}. These resources address embedded systems, digital design and computer architecture.

\textbf{Cybersecurity:} the rapidly evolving cybersecurity field is documented in The Cyber Security Body of Knowledge (CyBOK)~\cite{CyBOK2021} and Cybersecurity Curricula 2017~\cite{CSEC2017}, addressing security principles, risk management, cryptography, and secure system design across all computing domains.

\textbf{Data Science:} data science, as an emerging computing discipline, is currently under active development with Computing Competencies for Undergraduate Data Science Curricula~\cite{CCDSC2021} providing preliminary guidance that emphasizes statistical foundations, machine learning, data management, and ethical considerations in data-driven decision making.

While these discipline-specific BoKs provide essential foundations, their distributed nature across multiple documents and structures creates practical challenges for curriculum designers who must aggregate knowledge from diverse sources, reconcile overlapping areas, and ensure comprehensive coverage of competencies required in professional practice.

\subsection{Employment Context and Workforce Demands}\label{sec2.1}

The global demand for computing professionals continues to accelerate, creating significant workforce challenges across developed nations. According to the U.S. Bureau of Labor Statistics~\cite{adams2020_computing_dominate_stem,bls_emp_by_detailed_occupation}, the software development sector alone is projected to add approximately 29,000 new positions annually in the United States. It's represents only a portion of the broader computing employment landscape, which encompasses over 165,000 annual job openings when accounting for both newly created positions and workforce turnover due to retirements. Computing occupations are expected to dominate STEM job growth in the coming decade, underscoring the critical importance of robust computing education programs that can meet this expanding demand. Similar workforce challenges manifest in France, where the technology sector faces acute recruitment difficulties. According to France Travail's 2025 employment survey~\cite{francetravail_bmo2025}, the information and communication technology sector reports 71,040 recruitment projects, with 42.1\% classified as difficult recruitments due to insufficient candidate availability. The shortage becomes particularly acute for positions requiring master's-level qualifications, with 34,900 recruitment projects experiencing 57.4\% difficulty rates. This substantial gap between workforce demand and qualified candidate supply demonstrates the urgent need for effective computing education programs that produce graduates with both theoretical knowledge and practical competencies aligned with industry requirements.

\section{Ontology-driven competency model applied to software engineering programs}\label{sec3}

\begin{figure}[!htbp]
  \centering
  \includegraphics[width=0.9\linewidth]{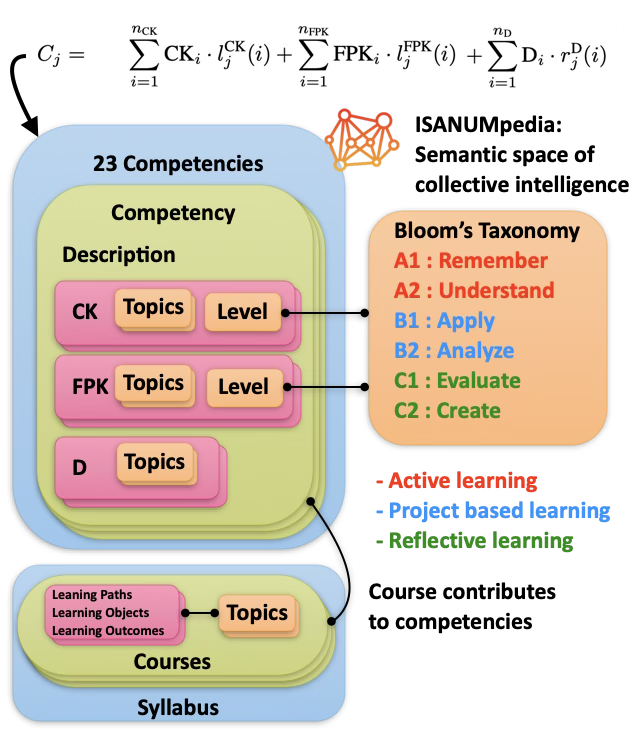}
  \caption{Competency mapping methodology}
  \Description{Representaition showing how course learning outcomes map to Computing Knowledge topics, which in turn map to program competencies, creating a traceable relationship between what is taught and what competencies students develop}
  \label{fig:competency-mapping}
\end{figure}

Our proposition aims to address the need to guide the various stakeholders involved in a training program in understanding the competencies expected at the end of the learning process. Indeed, it is necessary to share a common semantic model of the competencies expected by recruiters of future professionals, as well as by teachers when providing resources and learning activities, and finally by students so that they can be active participants in the process. Our proposal is based on the development of a shared model for the semantic representation of competencies, built on the BoK frameworks previously introduced, and includes the following concepts: competencies, disciplinary knowledge (for example, computing knowledge), professional knowledge, and skills. These concepts are interconnected through the identification of topics and levels of proficiency. Based on these elements, competencies can be defined through declarative ontological statements according to the following equation (see Figure~\ref{fig:competency-mapping})~\cite{ODCM}.

Once the expected competencies  of a training program have been constructed, and validated by potential employers, the training program can be developed in a precise and detailed manner through learning activities built around learning objects and learning paths. A course or learning path can be defined through a process based on a series of stages involving interaction with learning objects. Each learning object includes content and interactive activities, and in particular, assessments (diagnostic, formative, or summative) associated with expected learning outcomes, linked to topics and proficiency levels (i.e. bloom taxonomy). These objects and learning paths, created and validated by the teachers, can be used by the students, allowing them to follow unique or personalized paths, thus offering great flexibility. Providing explicit and transparent access to the competency model will allow students to understand the learning process, and enable self-learning, reflective-learning and self-assessment, making them active participants in the process. To illustrate the concrete benefits of this ontological competency model, the following section will describe its implementation in a collective intelligence space based on Semantic MediaWiki, implementing the previously introduced concepts and relationships, and allowing students, teachers, and representatives of the professional world to collaborate efficiently in the development of a computer engineering training program, which has been validated by the Commission des Titres d’Ingénieur (CTI), the only French authority accredited for this type of program.

\section{Case Study: ISANUM}\label{sec4}

To address the challenges of preparing computing professionals for digital transformation across multiple sectors, we propose a comprehensive five-year engineering degree program that integrates Bodies of Knowledge with competency-based education through a structured methodology. The ISANUM engineering degree program is designed as a specialized five-year curriculum that enables students to acquire domain-specific competencies from the first year while maintaining openness to research, innovation, and professional contexts. The program adopts active learning pedagogies, reflective learning and problem-based and project-based approaches, fostering connections with industry, addressing societal challenges, and promoting international perspectives. To measure what students learn, we use Computing Curricula 2020 (CC2020) recommendations~\cite{CC2020}, which identifies 34 Computing Knowledge elements (CK), 13 skills (FPK), and 11 dispositions (D) that students must develop. We measure knowledge using Bloom's taxonomy~\cite{krathwohl2002revision}. The taxonomy progresses through six cognitive levels: A1 (Remembering) focuses on recalling facts and basic concepts; A2 (Understanding) involves explaining ideas or concepts; B1 (Applying) requires using information in new situations; B2 (Analyzing) entails drawing connections among ideas; C1 (Evaluating) involves justifying a stand or decision; and C2 (Creating) requires producing new or original work. Dispositions are about attitude, like being proactive, self-directed, or inventive. We track these as required or not : Did the student need to show creativity in their project? To operationalize, we extracted and organized knowledge areas from curricula guidelines and Bodies of Knowledge. Our approach addresses the gap between Bodies of Knowledge and practical curriculum design through a competency-based framework. This section presents the ISANUM competency model, which translates the 494 knowledge topics from 34 Computing Knowledge areas into 23 concrete, assessable competencies organized into five thematic blocks. The ISANUM program structures its curriculum around five competency blocks that provide broad coverage of modern computing topics and opportunities for deep specialization. Each block addresses a critical domain of computing education identified in CC2020 and corresponds to specific industry workforce demands. Table~\ref{tab:competencies} presents the complete set of ISANUM competencies organized by block.

\begin{table}[!htbp]
\caption{ISANUM Competency Blocks and Competencies}
\label{tab:competencies}
\footnotesize
\begin{tabular}{@{}p{\linewidth}@{}}
\toprule
\textbf{ISANUM Competencies definition} \\
\midrule
\multicolumn{1}{@{}l}{\textbf{Block 1: Design and develop software solutions}} \\
\quad 1.1 Design software solutions meeting functional and non-functional requirements \\
\quad 1.2 Analyze complex problems and implement reliable algorithms \\
\quad 1.3 Apply software engineering methodologies and modeling frameworks \\
\quad 1.4 Develop solutions using appropriate programming paradigms \\
\quad 1.5 Conduct testing, validation and verification processes \\[0.5em]
\multicolumn{1}{@{}l}{\textbf{Block 2: Collect and process data, produce information}} \\
\quad 2.1 Design data structures and manage databases \\
\quad 2.2 Implement storage solutions for massive structured/unstructured data \\
\quad 2.3 Develop business intelligence and data warehouse solutions \\
\quad 2.4 Apply data mining and machine learning techniques \\
\quad 2.5 Develop knowledge representations and semantic web solutions \\[0.5em]
\multicolumn{1}{@{}l}{\textbf{Block 3: Design and develop IT infrastructures}} \\
\quad 3.1 Design IoT infrastructures with sensors and effectors \\
\quad 3.2 Configure network architectures for distributed systems \\
\quad 3.3 Program drivers and operating system components \\
\quad 3.4 Configure deployment environments for mobile and web applications \\
\quad 3.5 Develop cloud virtualization solutions \\[0.5em]
\multicolumn{1}{@{}l}{\textbf{Block 4: Design intelligent cyber-physical systems}} \\
\quad 4.1 Develop structural models for distributed CPS \\
\quad 4.2 Develop behavioral models for distributed CPS \\
\quad 4.3 Apply pattern-based design methodologies \\
\quad 4.4 Develop self-adaptive intelligent CPS \\
\quad 4.5 Integrate non-functional properties (safety, performance, energy) \\[0.5em]
\multicolumn{1}{@{}l}{\textbf{Block 5: Conduct multidisciplinary projects}} \\
\quad 5.1 Manage multidisciplinary research and innovation IT projects \\
\quad 5.2 Apply scientific methods to analyze complex systems \\
\quad 5.3 Lead international IT project teams with quality and ethics \\
\bottomrule
\end{tabular}
\end{table}

\begin{figure}[!htbp]
  \centering
  \includegraphics[width=0.85\linewidth]{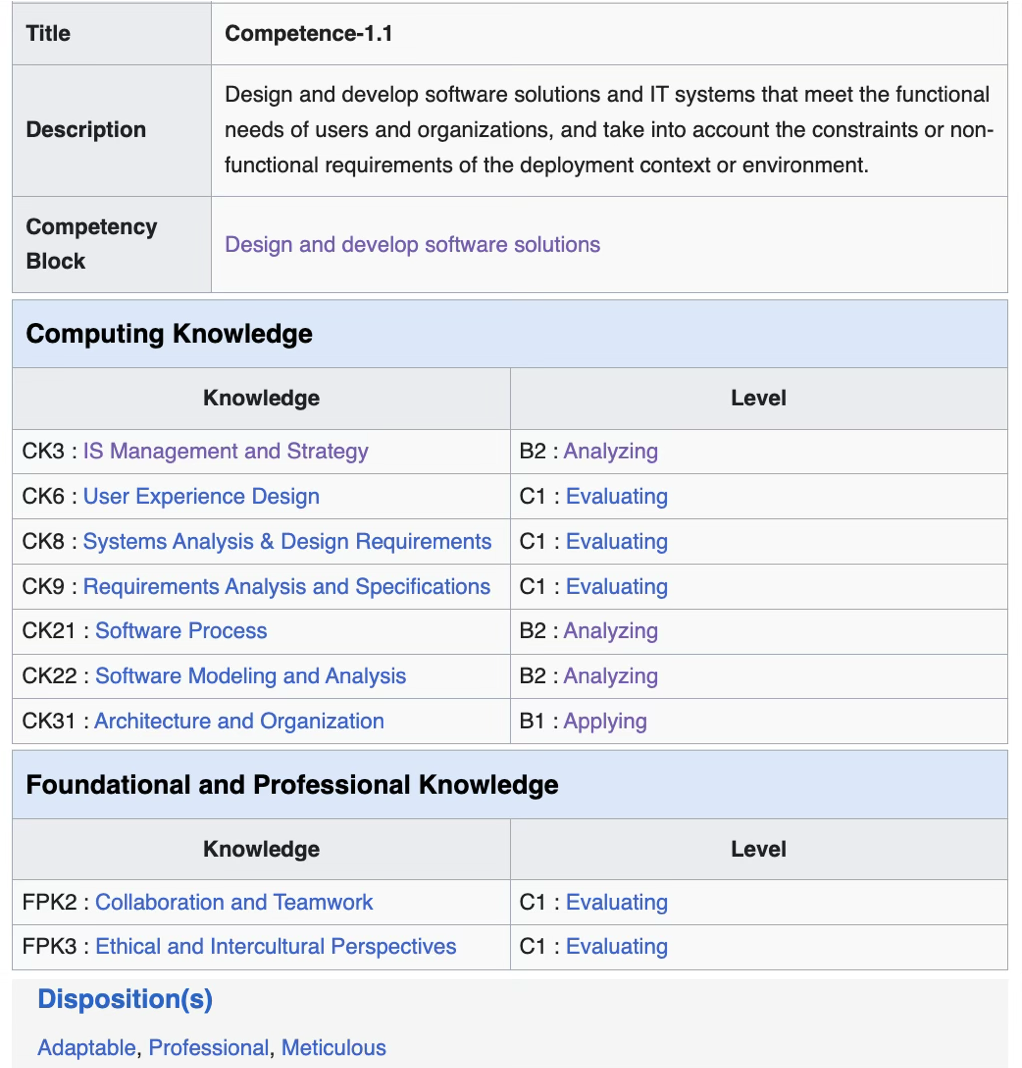}
  \caption{Structure of an ISANUM competency definition}
  \Description{Diagram showing how a single ISANUM competency integrates multiple Computing Knowledge topics mapped to Bloom's taxonomy levels, professional skills, and dispositional requirements}
  \label{fig:competency-definition}
\end{figure}

Each competency is formally defined as a structured combination of knowledge elements (CK), skills (FPK), and dispositions (D) that students must acquire at specified proficiency levels \cite{ODCM}. To maintain and evolve this competency model, we developed ISANUMpedia, a semantic MediaWiki platform built on a formal ontology. This wiki provides a semantic space for collective intelligence where all competency elements are represented as structured categories, properties and templates, enabling maintenance, version control, and collaborative refinement of the competency model. Figure~\ref{fig:competency-definition} illustrates the competency structure for one representative competency from the ISANUM program.

The mapping between BoK knowledge areas and competencies enables clear traceability throughout the curriculum. Each ISANUM competency explicitly lists which Computing Knowledge (CK), Foundational and Professional Knowledge (FPK), and Dispositions (D) students must master. When instructors design courses, they select topics to create their learning outcomes. Because each CK topic is linked to specific competencies, this creates an automatic connection: teaching a topic means contributing to the competencies that depend on it. This bidirectional mapping ensures that every course contributes measurably to program-level competency development, while allowing instructors flexibility in how they teach individual topics. For example, when a course teaches "Software Design" (CK area from SW Development category) and selects related topics, it automatically contributes to competency 1.1 "Design software solutions meeting functional and non-functional requirements" because this competency explicitly references Software Design knowledge topics in its definition. This approach ensures curriculum coherence while providing flexibility in course design. The five-year ISANUM program follows a progressive structure that develops both foundational knowledge and specialized expertise. Years 1-2 establish foundational competencies across all five blocks, ensuring students develop core computing knowledge and skills. Year 3 introduces mandatory alternance (work-study arrangements), where students alternate between academic terms and workplace immersion, applying their developing competencies in authentic professional contexts while continuing their coursework until the end. Years 4-5 enable specialization through three distinct pathways, each emphasizing different competency blocks while maintaining baseline proficiency across all areas:
\begin{itemize}
\item \textbf{Software Engineering specialization} emphasizes Blocks 1 and 5, deepening expertise in software development methodologies, quality assurance, and team-based project management for large-scale software systems.
\item \textbf{Data Engineering specialization} focuses on Block 2 and related infrastructure (Block 3), preparing students for roles in data science, business intelligence, and knowledge engineering across various application domains.
\item \textbf{Information Technology specialization} concentrates on Blocks 3 and 4, developing expertise in distributed systems, cloud computing, IoT infrastructure, and intelligent cyber-physical system integration.
\end{itemize}

The mandatory alternance component throughout years 3-5 ensures that students gain authentic workplace experience and develop professional competencies (particularly Block 5) alongside their technical specialization. This work-integrated learning model addresses the workforce preparation challenges identified in Section~\ref{sec2.1}, producing graduates with both theoretical depth and practical readiness for immediate professional contribution.

\section{Results}\label{sec5}

The ISANUM program enrolled its first cohort of 23 students in 2024-2025. The curriculum implements 60 ECTS credits per year, comprising approximately 600 hours of face-to-face instruction and 900 hours of self-paced learning activities. Students engage in reflective learning practices by developing individual e-portfolios that document their competency development. These e-portfolios leverage the ISANUMpedia ontology-driven competency model, enabling students to perform structured self-assessments aligned with the 23 program competencies. Professional partners are using these e-portfolios during recruitment for the mandatory work-study program, which begins in 2026-2027 and continues for three years. This integration of competency-based assessment with professional recruitment demonstrates the practical value of explicit competency frameworks in bridging academic preparation and workforce integration. Platform analytics from the first academic year (September 2024 - September 2025) show substantial engagement: 23 students created 931 achievement pages (averaging 8.47 competency-linked learning specifications each, 7,805 total) with 3,747 total revisions, indicating iterative reflection rather than one-time documentation.

\section{Conclusion and perspectives}\label{sec6}

This paper addresses a fundamental challenge in computing education: translating comprehensive Bodies of Knowledge into effective, competency-based curricula that prepare graduates for real-world digital transformation challenges. Through the ISANUM engineering program, we demonstrate a competency mapping methodology that bridges the gap between theoretical BoKs and practical workforce preparation. Our main contributions include: (1) a competency model and its practical implementation based on 23 clearly defined competencies organized into five blocks, each with explicit mappings to BoK knowledge areas; (2) detailed knowledge maps that translate 494 BoK topics from 34 Computing Knowledge areas into teachable content; (3) three specialized pathways (Software Engineering, Data Engineering, and Information Technology) that address digital transformation challenges across multiple sectors; and (4) mandatory alternance starting in year three, ensuring students develop both theoretical foundations and practical workplace competencies through work-integrated learning. This work provides computing educators worldwide with a replicable methodology for designing programs that connect Bodies of Knowledge to competency-based curricula, validated through the successful implementation of ISANUM. The semantic MediaWiki infrastructure (ISANUMpedia) supports collaborative maintenance and evolution of the competency model. Future work will focus on longitudinal evaluation of the ISANUM approach, assessing graduate outcomes, student and employer satisfaction, and the effectiveness of competency-based assessment through e-portfolios. However, maintaining curriculum alignment with rapidly evolving BoKs, standards, and industry requirements demands continuous integration efforts that impose significant workload on curriculum designers and educators. Additionally, future research should explore how AI agents can assist in monitoring BoK updates, suggesting competency mappings, identifying coverage gaps, and proposing curriculum adjustments, thereby enabling educators to focus on pedagogical decisions.


\bibliographystyle{ACM-Reference-Format}
\bibliography{main}










\end{document}